\documentclass[preprint,12pt]{aastex}
\usepackage{epsfig}
\usepackage{emulateapj5}
\usepackage{apjfonts}

\begin{document}
\submitted{Submitted to ApJ} 
\def\num1{${\cal D}_1$}
\def\mess{${{\theta_E}\over{\theta_1}}$}
\def\tetae{{$\tau_1 ({\rm days})$}}  
\def\br1br{${\cal R}_1^b ({\rm yr}^{-1})$}  
\def\r1d{${\cal R}_1^d ({\rm yr}^{-1})$}  
\def\rtb{{\bf ${\cal R}_{tot}^b ({\rm yr}^{-1})$}}  
\def\rtd{\bf{${\cal R}_{tot}^d ({\rm yr}^{-1})$}}   
\def\gc{globular cluster}
\def\kms{{\rm km/s}}
\def\rmeq#1{\eqno({\rm #1})}
\def\gl{gravitational lens} \def\gb{Galactic bulge}
\def\lc{light curve}
\def\ml{microlensing} \def\mo{monitor}
\def\pr{program} \def\mlmpr{\ml\ \mo\ \pr}
\def\ev{event}
\def\ex{expansion}
\def\fn{function} \def\ch{characteristic}
\def\bi{binary}  \def\bis{binaries}
\def\rd{Di\thinspace~Stefano}
\def\bl{binary lens} \def\dn{distribution}
\def\pop{population}
\def\ct{coefficient}
\def\cc{caustic crossing}
\def\cs{caustic structure}
\def\mag{magnification}
\def\pl{point lens}
\def\dm{dark matter}
\def\rd{Di\thinspace~Stefano}
\def\tots{track of the source}
\def\de{detection efficiency}
\def\det{detection}
\def\ob{observ}
\def\ol{observational}
\def\od{optical depth}
\def\ml{microlensing}
\def\mtm{monitoring team}
\def\mmtm{microlensing monitoring team}
\def\otm{observing team}
\def\mo{monitor}
\def\motm{microlensing observing team}
\def\los{line of sight}
\def\ev{event}
\def\by{binarity}
\def\ptb{perturb}
\def\sgf{significant}
\def\bis{binaries}
\def\sg{signature}
\def\bbl{binary-lens}
\def\kms{{\rm km/s}}
\def\rmeq#1{\eqno({\rm #1})}
\def\gl{gravitational lens} \def\gb{Galactic bulge}
\def\lc{light curve}
\def\ml{microlensing} \def\mo{monitor}
\def\pr{program} \def\mlmpr{\ml\ \mo\ \pr}
\def\ev{event}
\def\ex{expansion}
\def\fn{function} \def\ch{characteristic}
\def\bi{binary}  \def\bis{binaries}
\def\rd{Di\thinspace~Stefano}
\def\bl{binary lens} \def\dn{distribution}
\def\pop{population}
\def\ct{coefficient}
\def\cc{caustic crossing}
\def\cs{caustic structure}
\def\mag{magnification}
\def\ppl{point-lens}
\def\bl{blending}
\def\mage{magnification}
\def\fsse{finite-source-size-effects}
\def\cc{caustic crossing}
\def\mp{multiple peak}
\def\lcf{lightcone fluctuation}
\def\wdf{white dwarf}
\def\pn{planetary nebula}
\vskip -.8 true in
\title{
Mesolensing: I. High Probability Lensing Without Large Optical Depth}

\author{Rosanne Di\thinspace~Stefano}
\affil{Harvard-Smithsonian Center for Astrophysics, 60
Garden Street, Cambridge, MA 02138}
\affil{Department of Physics and Astronomy, Tufts
University, Medford, MA 02155}
\submitted{Submitted to ApJ, 22 April 2005} 
\def\gl{gravitational lensing}
\def\Gl{Gravitational lensing}
\def\ml{microlensing} 
\def\Ml{Microlensing} 
\def\Et{Einstein angle} 
\def\et{$\theta_E$} 
\def\Er{Einstein radius}
\def\ev{event}
\def\vb{variable}
\def\vy{variability}
\def\sg{signature} 
\def\asec{arcsecond}

\begin{abstract}
In a variety of astronomical situations, there is a relatively high
probability that a single isolated lens will produce a detectable event.
The high probability is caused by some combination of a large Einstein
angle, fast angular motion, and a dense background field. We refer
to high-probability lenses as mesolenses. The mesolensing regime is
complementary to the regime in which the optical depth is high, in that
it applies to isolated lenses instead of to dense lens fields. Mesolensing
is complementary to microlensing, because it is well suited to the study
of different lens populations, and also because different observing and
analysis strategies are required to optimize the discovery and study of
mesolenses. Planetary and stellar masses located within $1-2$ kpc are
examples of mesolenses. We show that their presence can be detected 
against a wide variety of background fields, using distinctive signatures 
of both time variability and spatial effects. Time signatures can be
identical to those of microlensing, but can also include baseline jitter, 
extreme apparent chromaticity, events well fit by lens models in 
which several independent sources are simultaneously lensed, 
and sequences of events. Spatial signatures include astrometric lensing 
of surface brightness fluctuations, as well as patterns of time 
variability that sweep across the background field as the lens moves. 
Wide-field monitoring programs, such as Pan-STARRS and LSST, are 
well-suited to the study of mesolensing.
In addition, high resolution observations of the region behind a
known mass traveling across a background field can use lensing effects
to measure the lens mass and study its multiplicity.
\end{abstract} 
\section{Mesolensing}

\Gl\ was first observed by measuring shifts, on the order of an arcsecond,
in the apparent positions
of distant bright sources, 
such as stars lensed by the Sun (Dyson, Eddington, \& 
Davidson 1920; Eddington et al.\, 1919)  or 
quasars lensed by intervening galaxy clusters 
(Walsh, 
Carswell, \& Weymann 1979). 
As the angular resolution of telescopes has improved,  
the spatial effects caused by lensing have helped to probe
the mass distributions of intervening large mass distributions.

Even when images cannot be resolved, time variability may provide
evidence that lensing has occurred.  
Lensing of quasars by stars in distant 
galaxies, e.g.,
 produces image separations on the order of microarcseconds,
generally too small to be resolved. 
Microlensing can be studied, however, by observing time variations in the
image fluxes (see Kochanek 2004 and references therein). In fact,
because the projected separation between stars in a distant galaxy can be
comparable to the size of their Einstein rings, the probability that an
individual source
is being lensed by an intervening star (i.e., the optical depth)
can be close to unity.
Furthermore, the high density of lenses is associated with caustic structures
that produce distinctive light curve features.     

The term ``microlensing" is also used to describe lensing by 
stellar-mass objects located, e.g., in the Halo, 
deflecting light from stars 
in neighboring galaxies. 
Although the image separations
are on the order of a milliarcsecond, they are typically
not resolved,
and it is  
time variability that
provides believable evidence of lensing. 
The probability that a given microlens in the Halo
is producing a detectable event are
typically small ($10^{-7}-10^{-5}$). 
The time signatures associated with the lensing of a single or
binary star by a point or binary lens are distinctive. They are difficult
to mimic in data sets that span several years (Paczy\'nski 1996).

In this paper we explore an intermediate (``meso") regime, in which
an individual isolated mass has a high probability 
of lensing a background source. The probability that a mass will serve
as a lens may be high because the Einstein angle is large, or
because the angular motion of the lens is large.
Large values of the Einstein angle and large angular
motions can also produce measurable spatial effects.
Signatures of mesolensing therefore include both
spatial and time effects.

The mesolens regime is complementary to the
regime in which the optical depth is high. In the latter case, the lens
density is large enough to ensure a high probability of lensing.
In the mesolensing case, the lensing probability is high because 
it is likely that, over a relatively short interval of time,
the Einstein ring of an individual lens  
will encompass 
the positions of sources bright enough 
to be detectably magnified.

While supermassive BHs in the centers 
of galaxies, or intermediate-mass BHs in our own Galaxy may serve as
high-probability lenses, the concrete examples we will focus on in
this paper (Di\thinspace Stefano 2005a)
are nearby stellar-mass objects
located within roughly a kpc of Earth, or planet-mass objects
somewhat closer to us.  
The main distinction between these nearby populations and the
classical microlens populations is that 
each mesolens has a higher probability of serving as a lens. 
The associated light curve profiles of ``events"
may be identical to or may be radically different from those
typically identified with microlensing. 
Spatial effects directly associated with lensing or with the motion
of the lens may also provide independent evidence of lensing.
In some cases the radiation from the lens may be directly detectable. 

\subsection{Nearby Objects as Mesolenses}

The properties of the local populations of WDs, NSs, and BHs remain
mysterious. We know nothing about isolated BHs (but see Mao et al.\, 2002;
Bennett et al.\, 2004), 
little about
isolated neutron stars (see, e,g., references in
Kaspi, Roberts, \& Harding 2004), 
and, although thousands of WDs have been
discovered (Nale{\. z}yty \& Madej 2004; Kleinman et al.\, 2004;
Luyten 1999; McCook, \& Sion 1999),   
 the distribution of WD masses
is not well established. The frequency of double degenerate
systems has not been directly measured, nor has the fraction of
neutron stars with planets, nor do we know whether WDs have planets.
Finally, recent surveys have discovered large populations of brown dwarfs
and low-mass stars, particularly in regions containing
young stars; complementary studies are important. 
Fortunately,
stars 
within roughly a kpc of Earth
can be discovered through their action as mesolenses, if they are positioned
in front of an appropriate stellar background.

If lensing can provide a systematic way to discover and study nearby dark 
and dim populations, this will be an important step for several
areas of astrophysics. 
We could, e.g.,  hope to determine the mass distribution 
for isolated neutron stars; to derive clues to constrain the
neutron star equation of state; determine the 
size and mass distribution of the
 population of isolated black holes. 

Although the obvious application of lensing studies is to
lenses that are dark or dim, all stars deflect light.
When light from a nearby nuclear-burning star can be masked, 
its action as a lens can be studied. This means that we can refine
mass estimates, and refine the trajectories of all nearby stars.
In principle, we can also 
determine the fraction with planets orbiting at distances
of a fraction of an AU to several AU (Di\thinspace Stefano 2005b).

\subsection{Mesolensing Monitoring Programs}  

The discovery of nearby dark and 
dim objects can be accomplished through monitoring programs. 
The most familiar type of monitoring is 
light curve monitoring, which has been used, e.g., to detect microlensing
events that might be caused by lenses located in the Galactic Halo. 
Mesolensing programs can benefit from light curve monitoring,
and also from monitoring that is sensitive to spatial effects.

\subsubsection{Light Curve Monitoring} 

The idea motivating previous and ongoing microlensing 
programs
is that monitoring
may discover lensing events providing evidence that
some fraction of Halo dark matter is in the form of
MAssive Compact Halo Objects (MACHOs). 
Crowded source fields are observed regularly, with cadences that
vary, but which may be on the order of once per day.
To mitigate the effects of blending, image differencing techniques are
implemented. Software programs use a set of criteria based on expected
event properties to
identify possible microlensing events.

While the existence of
MACHOs is still not well established, we know for certain that
there is a large population of dark and dim objects within 1 kpc 
of Earth. 
Some of these can also be found through monitoring programs.
Just as in the case of microlensing, the results are optimized through
the use of image differencing. When considering detectability issues
in this paper and its companion, we will therefore assume that 
image differencing is being implemented.  

In \S
 3 we compute the rate of well-isolated lensing events due
to nearby stars, and also consider the characteristics of
such events. We find that 
many mesolensing events are expected to be of shorter duration than
most microlensing events. This suggests that more frequent
monitoring than is typical for microlensing events may be helpful.
We also find that, although a minimum density of source stars is
required, 
mesolensing is more likely than microlensing to be detected
across a 
wider selection of background 
source fields.
Programs designed specifically for the study of nearby stars might therefore
choose somewhat different monitoring patterns than those used to date.

Nevertheless, data from
 previous and ongoing microlensing monitoring programs  
should contain events caused by nearby lenses. 
The rate computations of \S 3 are specialized
to specific background source fields in paper II.  
We will also derive the characteristic
features of events caused by nearby stars. We will find that, while
some light curves are identical with microlensing light curves,
others have different features. 

The challenges observers face in order to optimize the identification
of mesolensing events in existing data bases are to 
(1) design software to select
all events that could be due to lensing by nearby masses, 
(2) develop analysis tools to test the hypothesis that individual
events are caused by nearby lenses, (3) search existing data bases 
for possible counterparts to the lens, and (4) coordinate
additional mutliwavelength observations, when they seem to be necessary. 

By taking these steps, we can identify events in existing data bases
 caused by nearby lenses.  
This serves two purposes.
First it can significantly advance our knowledge of nearby
stellar remnants and low-mass dwarfs. 
Second, it clarifies the results of searches for MACHOs, by eliminating
contamination from events definitely {\it not} due to dark halo matter. 
As new monitoring programs come online, the cadence of observations
and background fields can be selected to yield further improvements
in our ability to detect and identify events due to nearby lenses.

\subsubsection{Monitoring Spatial Variations} 

Astrometric effects fall off slowly with distance from the lens.
They are inversely proportional to the distance, with ten percent effects
occurring at distances of up to $10\, \theta_E$.    
The Einstein rings associated with nearby lenses can be on the
order of hundredths of an arcsecond, instead
of milliarcseconds. This brings the detection of astrometric effects
associated with lensing into a regime where they are less difficult to
detect. In addition, the lens motion can be on the order of a tenth
of an arcsecond per year. This means that it may be possible to
detect lensing of source stars
located within a swath of width $20-40\, \theta_E$ ($\sim 0.3''$)  
and  length $\sim 10''$. At the distance to M31, this region would cover
tens of square parsecs and contain dozens of bright stars, and many more
stars too dim to be detectably lensed. Even if only a small fraction 
of stars are subject to astrometric effects significant enough to 
be detected, the combination of correlated changes following the track of
the lens in a regular way over time, may allow lensing to be identified
with a high degree of confidence. This possibility is discussed in \S 4. 

\subsection{Mesolensing by Known Masses}  

Beyond 
establishing some population properties,
whenever an event is established as 
having been caused by a nearby lens, additional study can be carried out.
This contrasts with the conventional view of microlensing, prevalent
when the programs began, that follow-up observations are unlikely to
teach us more about the nature of the lens.\footnote{There are now
several examples of follow-up that has been useful; the most 
 detailed information has been derived for MACHO-LMC-5, an event that
was caused by a nearby lens which has now been directly detected.}
Follow-up study of a mesolens 
can help to determine the nature and mass of the lens 
and to determine if it has companions.
 
The possibility of detailed study of a single object as it travels across a
dense field is not limited to objects discovered through their action as 
lenses.
A small subset of nearby stellar remnants
has already been discovered. 
In fact, we know of thousands of
possibly interesting mesolenses located within a kpc of the Earth.
These include pulsars, white dwarfs, L-dwarfs and other low-mass stars, 
brown dwarfs, and extrasolar planetary systems.
Any catalogued masses that lie in front of crowded fields
have relatively high probabilities of serving as lenses.       
In work in preparation,
we have begun the task of compiling a catalog of
high-probability lenses 
(Li 2003; Li, Pfahl, \& Di\,~Stefano 2005).
High-probability lenses can be monitored, and 
observations of mesolensing can measure lens masses and proper motions.
Under favorable circumstances, we can discover whether
some of these dim systems are binaries, and may also be able to extract a full
orbital solution {\it without using stellar spectra} (\rd\ \& Pfahl 2005).

\subsection{The Source Population}

Once the lens characteristics have been determined, 
it is also possible to learn more about the stellar density and luminosity 
function of the source field. This is also true of microlensing
events.
What mesolensing adds
is an addition to the event rate, providing larger numbers of events than
would otherwise be predicted and, 
in cases in which one lens generates a sequence of events, 
mesolensing allows individual small regions
of the source galaxy to be studied in more detail. 
Furthermore, as discussed in 
the companion paper (\rd\ 2005b), the distribution of event durations
for background fields (e.g., galaxies) with the same intrinsic
properties, provide a measure of the distance to the background
fields, while distributions of durations within
a single field map changing stellar densities and luminosity
functions across the field.   

\subsection{Overview of the Paper}
 
In \S 2 I discuss the spatial and time scales associated with mesolensing
and in \S 3, the basics of detection. 
The possibility of detecting spatial variations is discussed in    
\S 4. Section 5 is an overview
of the prospects  
for using 
mesolensing as a tool in astronomy.  

\section{Mesolensing Regimes}

\subsection{Large Einstein Rings} 

Let $M$ be the mass of the lens, $D_L$ the distance between the observer
and lens, $D_S$ the distance between the observer
and source plane, $x=D_L/D_S$. If the alignment
between the observer, lens, and source is perfect, the image
will be a ring with an angular size referred to as the
\Et, \et .  
\begin{eqnarray}
\theta_E & = & \Bigg[
    {{4 GM (1-x)}\over{c^2 D_L }}\Bigg]^{{1}\over{2}}\nonumber\\
         &   &\nonumber\\  
         & = & 0.01"\, \Bigg[{(1-x)\, \Big({{M}\over{1.4\, M_\odot}}\Big)\Big({{100\, {\rm pc}}\over{D_L}}\Big)}\Bigg]^{{1}\over{2}}\nonumber\\ 
\end{eqnarray}

\subsection{Source Density}

If
$N_S$ is the volume density of source stars, and $L_S$ is the 
radial extent of the source galaxy, then
\footnote{Let $z_S$ represent the
radial distance through the source galaxy as measured from the side
nearest us. In general $N_S=N(z_S),$ because the source
 density and absorption may both change change
over distances of tens or hundreds of pc.}
$\sigma_{S}$ is the number of stars per square arcsecond.
\begin{equation}
\sigma_{S}={{1.43 \times 10^4}\over{\Box''}}\,
           \Big({{N_S}\over{{\rm pc}^{-3}}}\Big)\,
           \Big({{L_S}\over{\rm kpc}}\Big)\,
           \Big({{D_S}\over{D_{M31}}}\Big)^2,
\end{equation} 
where $D_{M31},$ the distance to M31 is $780$ kpc
(Macri et al.\, 2001).

\subsection{Relative Velocities}    

Let $v_L$ represent the transverse velocity
of the lens relative to us; $\omega_L=v_L/D_L.$
\begin{equation}  
\omega_L = {{0.21''}\over{{\rm yr}}}\, 
\Big({{v_L}\over{100\, {\rm km s}^{-1}}}\Big)\,
\Big({{100 {\rm pc}}\over{D_L}}\Big)
\end{equation}  
The Einstein disks of some lenses traverse enough area in the course of
a year to encompass, at one time or another, the positions of many
stars.  
This increases the chance that the lens will pass close enough
to the position of a  bright star to produce a detectable event.
%as the Einstein disk spotlight travels across the face of the
%source plane, the added illumination of some source stars
%may become noticeable.     

If the proper motion of the lens is small, the dominant angular motion
may be associated with the motion of the Earth. Parallax produces an
angular motion of roughly
\begin{equation}  
\omega_p  =  {{0.04''\ g}\over{{\rm yr}}}\, 
\Big({{100 {\rm pc}}\over{D_L}}\Big)
= {{|{\bf g(t)}|}\over{5}}\, \omega_L\, 
\Big({{100\, {\rm km s}^{-1}}\over{v_L}}\Big).     
\end{equation} 
The average value of the magnitude of ${\bf g(t)}$ is 
determined by the position of the source relative to the ecliptic,
and is 
typically on the order of unity.
Both its value and direction relative to the independent transverse motion
of the lens  vary during the course of a year.

Parallax ensures that
the transverse motion of nearby lenses 
is large enough to allow the Einstein disk to cover the positions
of multiple stars during the course of a  year, increasing the 
likelihood of a detectable event. In addition, if a single lens
produces a sequence of events, or if the lens can be detected directly,
 we will be able to track its motion
across the source plane; when parallax contributes significantly
to the motion, the shape of the lens' track will bear  clear
and measurable signature of parallax. This will allow us to determine the 
distance to the lens.

The Sun's motion through the galaxy is associated with an angular
motion with respect to sources in a distant galaxy.
\begin{equation} 
\omega_S = {{6 \times 10^{-5}\, {\rm arcsec}}\over{{\rm yr}}} \,
           \Big({{v_S}\over{220 {\rm km\, s}^{-1}}}\Big)\,
	   \Big({{D_{M31}}\over{D_S}}\Big)      
\end{equation} 
This is supplemented (or diminished) by a comparable contribution from
motion of the source stars within their parent galaxy. Even though
this angular speed is relatively small, it can produce interesting
effects. For example, if a magnification of $100$ is needed to
detect an event for which $\theta_E = 0.01'',$ an angular speed
of $6.2 \times 10^{-5} {\rm arcsec\, yr}^{-1}$ would produce an
event lasting $1.7$ yrs. 
If the source stars are located near the center of a galaxy,
their speeds can be roughly $10$ times as large, and the local source
density can be $> 10^6$ pc$^{-3}$.

\subsection{Mesolensing}

The expressions above demonstrate that there are many situations in
which some combination of a large Einstein ring, a dense background field,
and large angular motion produces a situation in which
an individual mass is likely to lens its background in a way that can
be detected during the next year or decade.
When an individual lens has a high probability 
(close to unity) that
it is producing an event at any given time, or will produce
an event in the near future, we will refer to it as a
``mesolens". 
If the position of a mesolens is known, it can be
monitored in such a way as to increase the likelihood of detecting
its action as a lens, 
through some combination of light curve and astrometric study.  
For populations with few known individuals (such as isolated NSs and BHs),
monitoring programs can help to discover individual lenses.
As we will show below,
some features of the required
monitoring and analysis are the same as those 
presently used for microlensing. Others can be optimized by using different
strategies. 
Calculations and predictions must generally take into account
the presence of multiple sources in the field of view, more than one of
which may be significantly lensed at one time. 
 
\subsection{Stars Entering the Einstein Ring: Rate Per Lens}
 
We first define the event rate, ${\cal R}_1^E$ to be the rate at which source stars
enter the Einstein ring of an individual lens. The subscript ``$1$'' 
indicates that this is the rate associated with a single lens; the superscript
``E" indicates that this is the rate at which source stars enter the Einstein
ring of the lens.  
With this definition,
we can compute ${\cal R}_1^{E} = {\cal R}_1^E(M,v_L,D_L;N_S,L_S,D_S),$ the
rate of events associated with a single lens of mass $M$ and transverse
velocity $v_L,$ located a distance $D_L$ from us, 
if it is moving in front
of a population of source stars 
$D_S$ from us whose surface density is
characterized by $N_S$ and $L_S.$
\smallskip  
\begin{eqnarray}
{\cal R}_1^E & = &  2\, \theta_E\, \omega_L\, \sigma_S \nonumber\\
   &   & \nonumber\\  
   & = & {{60}\over{{\rm yr}}}\ \    
     \Big|{{\bf {v}_L}\over{100\, {\rm km s}^{-1}}} +  
          {{{\bf g(t)}}\over{5}}\Big|     \, \  
           \Big({{N_S}\over{{\rm pc}^{-3}}}\Big)\,
           \Big({{L_S}\over{\rm kpc}}\Big)\nonumber\\ 
       &   & \nonumber\\ 
       &   & \times \Big({{M}\over{1.4\, M_\odot}}\Big)^{{1}\over{2}} 
           \Big({{100\, {\rm pc}}\over{D_L}}\Big)^{{3}\over{2}}\nonumber\\ 
       &   & \nonumber\\  
       &   & \times \Big({{D_S}\over{D_{M31}}}\Big)^2
           \Big(1-x\Big)^{{1}\over{2}}\nonumber\\ 
\end{eqnarray}

The expression above includes the effects of parallax;
the vertical bars enclosing the vector sum represent the magnitude of
the vector.    
As the expression indicates, the overall
effect of parallax is an increase in path length, producing a
proportionately higher
rate.
As the relative contribution of the transverse component of the 
velocity decreases, parallax effects will significantly influence the
shape of the light curve.  
 An interesting limit is reached when the 
independent transverse speed is zero,
and all events are simply due to the Earth's motion;
events will repeat in a manner that depends on the position
of the lens.  
In fact, depending on the distribution of spatial velocities and
orientations, a small fraction of all lenses will have their transverse
motion dominated by parallax.  

Note that, for the parameter values used for normalization,
 ${\cal R}_1^E$ is larger than the number of Einstein diameters 
per year covered by the lens. This is because, at any given time,  
more than one source star can be within the Einstein ring. 
Yet, only a fraction of all of these lensed stars will be bright enough
that they can produce an event when the lens approaches at a distance of 
$\sim \theta_E.$ 

Astrometric effects can also be important, especially since the
size of shifts in image position fall off slowly with 
distance from the lens.
For example, the rate, ${\cal R}_1^{10\, E},$ 
at which stars enter a swath within $10\, \theta_E$ of the lens
is ten times larger than ${\cal R}_1^E,$ and shifts of $10\%$ in
the position of positive parity images occur at $\theta= 10\, \theta_E.$
Although in many cases astrometric shifts will be difficult or
impossible to measure, in others they can help to verify that light
curve effects are due to lensing. In still others, astrometry
may provide the first hint that lensing is occurring against the
background
of a distant stellar field.  
  
\subsection{Event Durations}

Define $\tau_E$ to be the time taken for the source to pass across
the Einstein ring of a lens.
\begin{eqnarray}
\tau_E &=&{{2\, \theta_E}\over{\omega}} \nonumber\\ 
       &=&35{\rm days}\, \Big({{M}\over{1.4\,  M_\odot}}\Big)^{{1}\over{2}} 
          \Big({{100\, {\rm km\, s^{-1}}}\over{v_T}}\Big)
          \Bigg[{{D_L}\over{100\, {\rm pc}}}\, (1-x)\Bigg]^{{1}\over{2}}      
\end{eqnarray}
For nearby lenses, the time duration increases with $D_L,$ reaching
$111$ days for a lens at $1$ kpc, when all of the other parameters
are held at the values chosen above. The value of $\tau_E$ continues to
increase until $D_L=D_S/2,$ and then decreases symmetrically.
Thus, if the distributions of lens masses and velocities within
the source population are comparable to those found within roughly
a kpc of Earth, the range of event durations due to nearby
lenses would be similar to that due to self-lensing of the source population.  

The Earth's motion can naturally affect the value of $\tau_E.$ 
The duration of short events will shrink or stretch according
to the instantaneous relative direction between the transverse
velocity, ${\bf {v}_L}$, and ${\bf g(t)}.$ For events with durations close a year,
the duration will not be much affected, but the effect of accelerations
is to alter the light curve
shape.

\subsection{Total Rate at which Stars Enter the Einstein Ring}

The {\it total rate} of mesolensing is the rate at which stars in the source galaxy enter
the Einstein rings associated with {\it all} intervening high-probability lenses.    
Its value 
depends on the spatial and velocity
distribution of lenses. Let $f(M,v_l,D_l)$ represent
this distribution, with
\begin{eqnarray}
n_L & = & \nonumber\\
   &   &4\, \pi\, \int_0^{D_L^{max}} dD_L D_L^2 \int_0^{v_l^{max}} dv_L 
\int_{M_{min}}^{M_{max}} dM\, f(M,v_L,D_L)\nonumber\\   
\end{eqnarray}
where $n_L$ is the number of lenses located within a distance
 $D_L^{max}$ of us.

We 
introduce $\Omega_{gal}$, the solid angle covered by the source galaxy, 
expressed in units of square degrees. 

Using the rate per lens, ${\cal R}_1^E$ from above,
the total rate for an individual galaxy is 

\begin{eqnarray} 
{\cal R}_{tot}^E &=& \nonumber\\
  & & \nonumber\\
& &{{4\, \pi}\over{4.125 \times 10^4}}\ 
{{\Omega_{gal}}\over{\Box^0}}\nonumber\\   
 &  & \times \int_0^{D_L^{max}} dD_L D_L^2  \nonumber\\
 &  & \times  \int_0^{v_l^{max}} dv_L \nonumber\\
 &  & \times \int_{M_{min}}^{M_{max}} dM\, \nonumber\\
 &  &\\
 &  & \times  f(M,v_L,D_L)\, {\cal R}_1^E(M,v_L,D_L;N,L,D_S) 
\end{eqnarray}

To derive some general results we make simplifying assumptions
about the form of $f(M,v_L,D_L)$. 
\begin{equation}
f(M,v_L,D_L) = \delta(M - \tilde M)\, 
\delta({\bf v_L}-{\bf {\tilde v}_L})\, N_L(D_L)  
\end{equation} 
The $\delta-$function form for the lens mass roughly 
corresponds to considering different types of lenses separately.
It makes sense, for example, to
consider WDs separately from NSs or BHs, since the spatial
density of lenses is different for each type of stellar remnant.
%\begin{eqnarray} 
%{\cal R}_{tot} & = & \nonumber\\
%& & {{1.7}\over{yr}}\, {{\Omega_{gal}}\over{^o}}\,  
%{{N_S}\over{{\rm pc}^{-3}}}\, {{L_S}\over{{kpc}}}\, 
%\Big({{D_S}\over{D_{M31}}}\Big)^2 \times \nonumber\\
%& &  
%\Big({{\tilde v}\over{100\, {\rm km/s}}}\Big)\,  
%\Big({{\tilde M}\over{1.4\, M_\odot}}\Big)^{{1}\over{2}}
%\times \nonumber\\
%& & 
%\int_{D_L^{min}}^{D_L^{max}} dD_L\, N_L(D_L)\, 
%\Big(1-{{D_L}\over{D_S}}\Big)^{{1}\over{2}}
%D_L^{{1}\over{2}}  
%\end{eqnarray}  

If we consider a fixed value for $N_L$, and consider nearby lenses, so that
$D_L<<D_S,$ then
\begin{eqnarray}
{\cal R}_{tot}^E & = &\nonumber\\ 
 &  & {{3.8 \times 10^5}\over{\rm yr}}\, {{\Omega_{gal}}\over{\Box^o}}\,
{{N_L}\over{{\rm pc}^{-3}}}\, 
{{N_S}\over{{\rm pc}^{-3}}}\, {{L_S}\over{{\rm kpc}}}\,
\Big({{D_S}\over{D_{M31}}}\Big)^2\times \nonumber\\
& & 
\Big|{{\bf {\tilde v}_L}\over{100\, {\rm km/s}}} + {{{\bf g(t)}}\over{5}}\Big|\,
\Big({{\tilde M}\over{1.4\, M_\odot}}\Big)^{{1}\over{2}}
\Big({{D_L^{max}}\over{{\rm kpc}}}\Big)^{{3}\over{2}}
\end{eqnarray}  
The rate of detectable events is smaller. 

\section{Detecting Mesolensing}

When a mass lies in front of a dense source field,
a pattern of perturbations in the light distribution
is created. The surface brightness 
fluctuations that would have been observed had
the deflecting mass not been there, are redistributed. 
When the mass travels across the face of the field, 
the pattern of perturbations changes and moves, introducing
time variations that %, like the wake of a boat
%moving in water, 
travel across the source field.

Some characteristics of the expected perturbations are the
following. (1) They should move in a smooth progression along
a well-defined path. (2) For 
each region in which perturbations are observed,
there should be a baseline to which the light received 
per unit time returns. Time 
variability occurring before or
 after 
the lens passes should be
be traceable to the ordinary variability
of stars in the region. Other observed variability may be caused
by alignment issues an/or by the relatively slow motion of stars
in the source galaxy into and out of the region.

Because lensing preserves surface brightness,
we should be able to verify that the surface brightness is constant when
(3) 
averaged over space at any one time, 
and (4) averaged over time in any one region.

If angular resolution on angular scales comparable to
$\theta_E$ is available, 
the distribution of surface density fluctuations can be at least roughly
tested against lensing model predictions. As a point lens moves across
the field, a small disk with relatively steady emission will surround
the lens position, traveling with the lens in a way
that can be compared to the calm in the eye of a hurricane. A region of 
more pronounced fluctuations will trace the moving Einstein ring.  
Even if the angular resolution is
significantly worse,
however,
high-amplification events are expected and can be
fit by simple physical models to
verify the hypothesis that the observed perturbations 
are due to mesolensing.    
 
\subsection{The Significant Angular Scales} 

Four angular scales determine the characteristics of the signatures of
lensing. 
The first is the Einstein angle, given by Eq.\, (1).
The value of $\theta_E$ depends on the  
mass and position of the lens. For nearby lenses, $x=D_L/D_S<<1,$
so the distance to the background source field doesn't play
a significant role.  

The second scale is $\theta_1,$ the angle
which encloses, on average, a single source star. The value of
$\theta_1$ is determined by the surface density of source stars.
\begin{equation} 
\theta_1 =
\sqrt{{{1}\over{\pi\, \sigma}}}= 
 4.7 \times 10^{-3}\ '' \, \Bigg( 
       {{10^3\, {\rm pc}^{-3}}\over{N_S}} 
       {{{\rm kpc}}\over{L_S}}\Bigg)^{{1}\over{2}}
       \Big({{D_{M31}}\over{D_S}}\Big)  
\end{equation}

The third scale  
is the angle being monitored by our detectors, $\theta_{mon}.$
The resolution of the 
telescope being used for monitoring places a lower limit on 
$\theta_{mon}.$
The number of stars monitored is 
\begin{equation} 
N_{mon}=\sigma\, \theta_{mon}^2, 
\end{equation}  
where we assume that the region monitored is a square with 
side $\theta_{mon}.$ 

The fourth relevant scale is set by the value of $\theta_{b,i},$
the angle of closest approach between lens and source required in order
for the source by itself to be sufficiently magnified
to produce a detectable deviation above baseline.   
If the lensed source star is bright, contributing 
most of the light from the monitored region, 
the value of $\theta_{b,i}$
can be equal to $\theta_E$, or even somewhat larger.
In general, however, $\theta_{b,i}<\theta_E$. 
It is often convenient to use the dimensionless quantity 
$b_i=\theta_{b,i}/\theta_E.$   

\subsection{The Monitored Field}

The total baseline luminosity is
\begin{equation} 
L_0 = \sum_{i=1}^{N_{mon}} L_i,
\end{equation}
where 
$L_i$ is proportional to the number of photons 
we receive from the $i\, $th star in the monitored region 
if there is no lensing. 
In the case of mesolensing, more than one star at a time
 may be significantly
magnified.
The total luminosity from the monitored region at time $t$ is  
therefore 
\begin{equation} 
L(t) = \sum_{i=i}^{N_{images}} A_i(t) L_i= A_T(t)\, L_0 = \Bigg(1+f_T(t)\Bigg) L_0 
\end{equation} 
$N_{images}$ is the number of images in the monitored region.
$A_T(t)$ is the total magnification, and $f_T(t)$
is the fractional change in photons received from the monitored field.

Detectability is defined by a threshold value of $f_T(t)$.  
The required threshold depends on the physical situation.
If the source density is high enough that hundreds of stars
lie in the monitored region, then, when we return to the
view the same area time after time, small misalignments will
mean that a handful of stars could be added in some views, but
not others. This can lead to apparent variations on the order 
of a percent or more. Stellar variability can produce comparable effects. 
Thus we expect that $f_T(t)$ must typically be at least $0.1,$
in line with the characteristics of published microlensing
candidates observed in M31 
(e.g., Paulin-Henriksson et al. 2003). 
In fact, the successful observations of
lensing in  M31 illustrate that  
it is possible to meet the challenge of  
detecting a lensing event or events against 
background light, which may itself be 
fluctuating because of ordinary stellar variability.
These observations have used image differencing, which    
has become standard procedure in microlensing
observations, and have 
identified several good microlensing candidates
(Paulin-Henriksson et al.\, 2003; de~Jong et al.\, 2004).

The value of $N_{images}$ is different 
from that of $N_{mon},$ the number of 
source stars with positions in the monitored 
region. Every star in the source galaxy contributes a ``negative
parity" image; this image is located within the Einstein ring and
is highly demagnified ($A << 1$), producing arbitrarily small
luminosities as the transverse distance between the lens and source
increases.   
The images that are the most highly magnified are 
are those associated with stars close to the projected 
position of the lens. These high-magnification images  
are located near the critical curves. If the lens is a point mass,
the critical curve is a circle with angular radius equal to $\theta_E.$
%If the lens is a multiple, the dimensions of the critical curves are
%also comparable to $\theta_E,$ but they can be somewhat larger and can also
%exhibit significant structure.

In the extreme case in which the source field is so dense that
it can be viewed as a continuous screen of light, lensing cannot
be detected. The magnification of sources very near to the lens
or to any associated caustic structures will
enhance the amount of light received from these stars, but the 
associated increase in area  will push stars out of the monitored region,
eliminating an equal amount of light.
Surface density
fluctuations can, however, be detectably lensed.
Mesolensing therefore spans the regime between the relatively
straightforward detection of lensing of a 
single star, to the undetectable lensing of a uniform region. 

\subsection{Detectable Events}

We will focus on the case in which an observable deviation from baseline
is due to the lensing of a small number of stars. 
The same general considerations are necessary to study 
``pixel lensing" (see, e.g., Gould 1996).  
Because we are monitoring a region larger than the Einstein ring, all of the 
positive and negative parity images of stars close to the lens are 
located in the monitored region. 
Since, by assumption, these play the major role, we  
can compute the main features of the expected signatures by considering
only stars located in the monitored region, and including the
magnification
of only the $n$ stars closest to the lens at a given time.
%Because we are generally monitoring a region with linear dimensions
%larger than $\theta_E,$ we 
%assume that the positive and negative parity images of each source are also
%in the monitored region.   
\begin{equation} 
L(t) = \sum_{i=1}^n A_i(t) L_i +  \sum_{i=n+1}^N L_i 
\end{equation} 
Defining $f_i$ to be the fractional change in light received 
from the $i$th source,  
\begin{equation} 
\sum_{i=1}^n f_i(t) L_i  = \Big[f_T(t)\Big]\, L_0.     
\end{equation}

\subsubsection{Approximate Rate}

For large magnification events, $f_i \sim A_i \sim 1/b_i$.
Consider the case in which the dominant contribution to the magnification
comes from a single star, with luminosity $L_i.$
In order for the total luminosity to increase by an overall
factor $f_T,$ this star must come within 
\begin{equation} 
b_i = {{L_i}\over{f_T L_0}}, 
\end{equation}
of the lens.
The size of the monitored region must be large enough that 
$L_0$ is not zero, so that the value of $b_i$ is well defined.

$L_0$ can be written as a product of the average luminosity
of source stars in the monitored region,  
$\langle L \rangle,$ times the number, $N_{mon}$ of
stars in the same region.  
Thus
\begin{equation}
{b_i} = {{L_i}\over{\langle L \rangle}}\, {{1}\over{f_T N_{mon}}}
\end{equation}
The rate of detectable single-lens events is 
\begin{equation} 
{\cal R}_1^{detect}={{2\, \theta_E\, \omega\, \sigma}\over{f_T N_{mon}}}\, 
        \int_{L_{min}}^{L_{max}}dL\, f(L)\, {{L}\over{\langle L \rangle}},   
\end{equation} 
where $\int dL\, f(L)\, L = {\langle L \rangle},$ and $L_{min}$ and $L_{max}$
are, respectively, the minimum and maximum source luminosities.
Note that this applies to the track of a single lens. If the track 
is long and covers a wide enough range of luminosities that the
integral above is approximately unity, then 	 
\smallskip

\begin{eqnarray} 
{\cal R}_{1}^{detect} & = & {{2\, \theta_E\, \omega}
     \over{f_T\, \theta_{mon}^2}}\nonumber\\   
             &   & \nonumber\\  
             &   & \nonumber\\  
             & = & {{0.042}\over{{\rm yr}}}\, \Big({{0.1}\over{f_T}}\Big)\, 
                   \Big({{1''}\over{\theta_{mon}}}\Big)^2\, 
\Big|{{{\bf {\hat v}_L}}\over{100 {\rm km s^{-1}}}} + 
{{{\bf \hat g}}\over{5}}\Big|\nonumber\\
             &   & \nonumber\\
             &   & \times  
                   \Big({{M}\over{1.4\, M_\odot}}\Big)^{{1}\over{2}}
	           \Big({{100 {\rm pc}}\over{D_L}}\Big)^{{3}\over{2}}
                   (1-x)^{{1}\over{2}} \nonumber\\   
\end{eqnarray} 
If we know the position of a lens, such as a cool dwarf star or a WD,
the chance of detecting a lensing event can be increased by 
monitoring a smaller area or by using very sensitive photometry.  

The total rate of detectable events can now be written as follows,
where we assume that $x << 1.$ 
\begin{eqnarray}
{\cal R}_{tot}^{detect} & = &\nonumber\\
 &  & {{270}\over{\rm yr}}\, {{\Omega_{gal}}\over{\Box^o}}\,
\Big({{0.1}\over{f_T}}\Big)\,
                   \Big({{1''}\over{\theta_{mon}}}\Big)^2\nonumber\\
	     &   & \Big({{N_L}\over{1 {\rm pc}^{-3}}}\Big)
\Big|{{{\bf {\hat v}_L}}\over{100 {\rm km s^{-1}}}} + 
{{{\bf \hat g}}\over{5}}\Big|\nonumber\\
             &   & \nonumber\\
             &   & \times 
                   \Big({{M}\over{1.4\, M_\odot}}\Big)^{{1}\over{2}}
\Big({{D_L^{max}}\over{{\rm kpc}}}\Big)^{{3}\over{2}}
\end{eqnarray}
Note that the density of most potential lenses is significantly
smaller than $1$ pc$^{-3}.$ For WDs, e.g.,
the value os roughly $10^{-1}$ pc$^{-3}$   
(Holmberg, Oswalt, \& Sion 2002; Kawka, Vennes, \& Thorstensen 2004;
Liebert, Bergeron,
\& Holberg 2005); the rate per year per square degree  
is on the order of a few per year.  
The key to observing a large number of events per year is to
survey a large area of the sky. 

\subsection{Background Independence}  

The above expressions for the rate of {\it detectable} events are
remarkable because they have no direct dependence on the
source density. This is because, although a higher source density
promotes a higher event rate, the required
distance of closest approach is decreased by increasing source density,
and the two effects cancel. 
Nor is there direct dependence on the distance of the source field, $D_S.$ 
The implication is that  
virtually any  mass 
that is either fast-moving or has a large Einstein ring can
be detected and studied through its action as a lens of distant
objects. 

Thus, the variety of possible background fields and lensable
sources is incredibly diverse. Galaxy halos can be considered,
as can the intracluster media of galaxy clusters. 
Indeed, the expressions for the rates given above
assure that mesolensing will be observed by various wide-angle
surveys, such as Pan-STARRS (Chambers et al.\, 2004)  and 
LSST (Stubbs et al.\, 2004), now being planned.
The ultimate background field will be the cosmic microwave background,
when fluctuations on small scales are detectable.
For cases in which the surface density of stars is prohibitively high,
such as in distant galaxies or in the dense nuclear regions of
nearby galaxies, 
ionization nebulae or even X-ray sources can be considered
as potentially lensed sources of light, instead
of stars.   

\subsubsection{Event Characteristics}

Perhaps the most important measurable characteristic of most events is the
duration above baseline, $\tau_i,$ equal to $b_i \times \tau_E.$ 
The average duration is $\tau_{avg} = \langle b \rangle \, \tau_E.$
The value of $\langle b \rangle$ is $(f_T N_{mon})^{-1}$. 
For $f_T \sim 0.1$ and $N_{mon} \sim 100,$ $\langle b \rangle \sim 0.1$.
Given the shape of typical stellar luminosity functions, a small number of
high-$L$ stars have luminosities greater than the average; the lensing
of these sources will produce events with durations greater than $\tau_{avg.}$.
Typical events will be of shorter duration. For a fixed value of 
$v_L$ and a fixed value of $D_L,$ the distribution
of durations will mirror the luminosity function. 
Because, however, the volume containing lenses increases  with the third power of
$D_L,$ the overall distribution of event durations will be markedly 
less skewed toward
short-duration events than the luminosity  
function is toward low-$L$ stars. 

It is nevertheless interesting that, 
because the integral over luminosities extends to the
least luminous stars in the source field, 
events with durations several orders of magnitude smaller than the 
average could occur. What determines the shortest duration observable?
Analogies with microlensing may suggest that it is finite-source size effects.
However, in mesolensing, the projected linear dimensions of the Einstein rings tend
to be large. ($0.01''$ corresponds to $\sim 8000$ AU   at the distance to M31.)
The limit typically comes, instead, from signal to noise (S/N) requirements.
If, e.g., a low-$L$ star is magnified, the value of $b_i$ required for
a detectable event may be so small, that $\tau_i$ is a minute or less.
Even if we happened to be observing the location at which the event is 
occurring, the value of $S/N$ integrated over an exposure may not be
large enough to allow the event to be detected. For any given observational
set-up, 
there is therefore an effective cut-off
of the distribution of time durations. Nevertheless,
depending on value of $\tau_E$ and the luminosity distribution
of the source field, some events with durations < 1 day
 may be detectable.

\subsubsection{Corrections}

Several approximations have been used. 

\smallskip
 
\noindent(1) We have assumed that we are monitoring a region in which 
there are stars that can be detectably lensed. This implies first
of all that  the baseline
luminosity should not be zero. Second, there should be some
value of $b_i$ for which an event could, in principle, be detected.
All of the unlensed stars should not be so dim, e.g., that the small values 
of $b_i$ required, produce events too short to be detected.

\smallskip
 
\noindent(2) We have assumed that the deviation is due to 
the lensing of a single source. While this tends to be the case for
the highest magnification regions of the highest magnification
events, the simultaneous 
lensing of several stars contributes to many events.
Multiple source events are generally more complex, and have
higher  duty  cycles. They are discussed in \S 3.5.

\smallskip
 
\noindent(3) We specifically considered events with
such high magnifications that $b \sim 1/A$.
This is a good approximation for $A> 2.$ 
We should therefore consider separately
 distances of closest approach larger than roughly $0.5.$
The only stars that can be detectably magnified for such
large values of $b$ are those with
\begin{equation}
L_i > L_{low} = {{f_T L(0)}\over{{{b^2 + 2}\over{b\, \sqrt{b^2 +4}}}-1}} 
\end{equation}
The monitored region  can contain at most $n$ such stars, where $n$ is small. 
For example,
with $f_T = 0.1,$ and $b \sim 0.5,$ $L_{low} > 0.1\, L(0);$
ten or fewer stars can satisfy this condition. 
To derive an exact expression for ${\cal R}_1^{detect}$, we
must truncate the luminosity integral in Eq.\, 21 at $L_{max}=L_{low}$,
and add a term for the contribution of more luminous stars.
The new term is the integral over 
luminosities of $2\, b_i\, \theta_E\, \omega_L\, \sigma,$  
where the value of $b_i$ will range from $0.5$ to a maximum that
could be as high as $\sim 1.7$, and the value of $\sigma$ 
for these bright stars is $n/N_{mon}$ times the value for all stars.
Thus, the functional form of this new term is the
same as that of ${\cal R}_1^{detect}$ derived above,
and, in addition, its value will be close to the value given 
above. We will therefore continue to use the estimates given in Eqs.\, 22 and 23.
Note, however, that, depending on the luminosity
function these estimates can be very close to the rates 
for bright stars alone.  

\smallskip

\noindent(4) We have averaged over the source luminosity, which
is equivalent to assuming that the portion of the track we are
considering is long enough to well-sample the distribution of
source luminosities. While this may be appropriate for ${\cal R}_{tot}^{detect},$
because many lenses are included, the value of ${\cal R}_1^{detect}$    
for a specific lens depends on the details of the source field behind
that lens. If, therefore, an individual lens is being monitored,
both the sampling strategy and predictions can be designed to 
suit that specific case and to improve the chances of event detection.  
By using better resolution and more sensitive photometry the rate of events
that can be well fit by lensing models  
can be increased by two orders of magnitude. In addition, measuring
spatial effects such as a sequence of perturbations that track the lens,
or centroid shifts due to lensing, can provide supplementary
information that verifies that lensing is occurring and 
helps to constrain models.

\subsection{More Complex Perturbations}

Above we considered the limit in which there are
well-separated events due to lensing of a single
source star. In some cases, however, 
a sequence of events (i.e., a set of ``repeating events"), 
 or even the simultaneous detectable lensing of
several sources is expected.  
These regimes are delineated by the value of $\theta_1$
and its relationship to $\theta_{mon}$ and $\theta_b.$

\noindent{\sl Ultra-low density:} When the monitored regions contain,
on average, less than a single star, we are in the ultra-low-density
regime: $\theta_1 > \theta_{mon}$.  
\begin{equation} 
\sigma_S < {{1}\over{\pi\, \theta_{mon}^2}} 
\end{equation} 
In this regime, events involving the simultaneous lensing
of more than one star are rare. In addition, blending of light
from the lensed source with other monitored sources will not
typically be important. The majority of light curves 
will be generated by individual source systems.  
Data analysis can proceed as it
would in the case of individual microlensing events. 
The high probability nature of the lensing is expressed 
only through
a high event rate, most likely caused by a fast-moving lens.   

\subsubsection{Low, Medium, and High Density}

Consider the case in which $\theta_1 < \theta_{mon}.$ 
Then, the condition that the time between events
($\sim 1/{\cal R}_1$), is significantly 
larger than typical event durations
($2\, \theta_b/\omega$), is 
\begin{equation} 
4\, \theta_b^2 < \pi\, \theta_1^2 
\end{equation} 
In other words, if we consider a disk of radius $\theta_b,$ 
the probability that it is occupied by a 
source star must be small. 

The part of the parameter space we will refer to as the {\it low density}  
regime is  
$\theta_b<< \theta_1.$ In the low-density regime, events
are  distinct and well-separated from each other. 
They are unlikely to repeat, and will simply be blended versions
of Pacz\'nski light curves. The amount of blending is determined by the 
ratio $\theta_1/\theta_{mon}.$ The duty cycle is low.  
The value of $\theta_b/\theta_1$ in the top left panel of Figure 1
is $0.018,$ depicting the low density regime.  

As the value of  $\theta_b/\theta_1$ increases, 
the duty cycle also increases, and there is a
higher probability of detectable ``repeats" in the course of a year.
The lower panels of Figure 1 illustrate the
{\it medium density} regime, with $\theta_b/\theta_1=0.09$ on the lower left,
and $0.18$ on the lower right.
If, as in the figure, the increase in $\theta_b/\theta_1$ is due to an
increase in the value of $\theta_b,$ then event durations also increase
in this regime.

As the value of  $\theta_b/\theta_1$ increases toward unity,
the circles of detectability associated with different  lenses
begin to merge. There are, therefore, not only repeating
events, but events of even longer duration. These latter
are due to the simultaneous lensing of several source stars.
  The duty cycle of activity
is large in the {\it high density} regime, shown in the upper right
panel of Figure 1, where $\theta_b/\theta_1=0.54.$        

Note that the terms low-, medium-, and high-density refer to the
values of the ratio $\theta_b/\theta_1,$ and not to the value of 
$\theta_1$ itself, which characterizes the surface density of source stars.
Figure 1 illustrates that, even with a constant surface density, the
changes in $\theta_b$ that span a factor of $10$ can span the
low- to high-density regimes. 

Table 1 applies to the same situation, illustrating how the duty cycle and
event characteristics change with increasing values of $\theta_b/\theta_1.$

In more realistic situations, the source stars will span a large range of
luminosities. The circle of observability around different stars
will have radii equal to $\theta_{b_i} \sim L_i/\langle L \rangle.$
The bright stars will therefore have the highest probability of
being part of a sequence of events; events involving bright stars
are the events most likely to deviate from the simple Paczy\'nski
form.  

\subsubsection{Model Fits}

By focusing on regions of the light curve with
high total magnification, we can effectively limit 
consideration to regions with arbitrarily small
values of $\theta_b$. This corresponds to 
entering the low-density regime, where well-separated single-source
models provide good fits. As we include lower-magnification
portions of the light curve, the appropriate models
depend on the intrinsic source density. If it is high, 
then acceptable fits may require multiple-source models.  
This is demonstrated in the companion paper.

\section{Spatial Effects}
During lensing there are shifts in the number of images, their
positions, and in the surface area
of each.
The light curve calculations carried out for the simulations in paper II
include all of
these effects. The high-magnification portions of the light curves
are, however, dominated by
changes in area of individual images.
These magnification effects fall off rapidly with distance.
\begin{equation}
A(u)-1 \sim {{1}\over{u^4}}
\end{equation}

Astrometric effects fall off less rapidly ($\sim {{1}\over{u}}$), 
 hence can in principle
be detected in a larger area around the lens. With significant
deviations occurring over a larger
area, they also apply to a larger number of source stars.
In Figure 2, I show the changes in image positions of stars 
in a region that is $20\, \theta_E$ on a side, centered on the
lens position. The source density is $1/\theta_E^2$.
The source
positions (black crosses) were generated randomly. 
Red crosses mark the positions of the
positive parity images.% (see the appendix). 
If $u$ is the
projected
distance between the source and lens, and $u_+$ s the
projected
distance between the positive parity image and lens,
then
\begin{equation}
d_+ = u_+ - u \sim {{1}\over{u}}
\end{equation}

The overall effect, which can be seen in Figure 2, is a differential
stretching of the locations of positive partity
images.
Equally important, the location of the lens is moving, so that
each region traversed by the lens
 experiences the stretching and then relaxes back to its original shape,
while the perturbation moves on to the next region.

The detectability of astrometric effects has been studied in a number of
different situations.
The case of a large, static Einstein ring projected against M31
and observed by {\it HST} was considered by Turner, Wardle,
\& Schneider (1990).  
Pattern magnifications were considered by Saslaw,
Narasimha, \& Chitre (1985).
On the other extreme, astrometric effects acting
on individual stars  
have also been considered (see, e.g.,  
Pacz\'ynski 1996b; Dominik \& Sahu 2000;  
Jiang et al.\, 2004). 

One significant feature of mesolensing by nearby stars 
is that the size of the Einstein rings
tend to be larger than in microlensing. 
This means that larger astrometric shifts are expected, and
can be more easily measured. Another important characteristic
 is that 
angular speeds tend to be high. Against a dense background field,
many stars may therefore 
come within roughly $10-20\, \theta_E$ of the lens
during a time interval of years. 
The large numbers of shifts expected, 
in a population of stars likely to exhibit surface density fluctuations,
mean that lensing effects can in principle be observed
with a high level of confidence. Estimates are made by
Propotopapas et al. (2006).

 In many mesolensing situations, it will not be possible to detect
for an individual star, the apparent image motion caused by lensing. 
Nor will the Einstein radius generally be large enough to resolve the
Einstein ring and study the sorts of effects considered by
Turner, Wardle,
\& Schneider (1990).
Nevertheless, the 
number of stars per year that come within $10\, \theta_E$ of the  lens
is ten times the rate ${\cal R}_1^E.$ The number of stars in a 
$20\, \theta_E$ swath covered by an individual lens per year
may therefore be large enough that it encloses a significant
 number of surface
brightness fluctuations.
(In some cases, nebulae may be included as well.) 
It may be possible to detect 
slight deviations in the shapes and positions
of these as the lens approaches and then departs.
Because the lensing model has few free parameters, and there
may be many individual motions to study, there should be cases
in which it is possible to verify the lensing model through studying
the time and spatial evolution of a large number of slight shifts.
Such studies would have much in common with
studies of weak lensing in the cosmological context, although 
time variability of the signal would provide additional useful
information.

\section{Conclusion} 
\subsection{Summary}

Mesolensing is high-probability lensing.
The high probability is associated with
some combination of  a large Einstein ring and fast angular motion.
Under a set of simple assumptions, 
the rate of events in which the total magnification 
of a monitored field exceeds any
given value is independent of the source density. This means that,
in principle, the entire sky is a suitable background in which to
search for evidence of mesolensing.
If, e.g., we could discern small-scale variations in the
microwave background (due, possibly, to the
Sunyaev-Zeldovich effect [see, e.g., Springel, White, \& Hernquist 2001
and references therein]), the entire sky could be a background against
which we can discover and study nearby stars.
For now, many stellar backgrounds can be used.

Although the event rate is
background-independent, the event characteristics are not. 
In more dense backgrounds, event durations are shorter, and
the duty cycle and incidence of simultaneous multi-source lensing
can each be larger. This means that, for identical source fields located
at different distances from us, event characteristics can be used 
as a distance indicator. This method to measure distances uses time variations
in a way that is analogous to the use of surface brightness
fluctuations (Tonry \& Schneider 1988;
see also references in Mei et al.\, 2005).    
Furthermore, for sources  
at a fixed distance, the event durations and incidence of multiple-source
events, map the source density.

The first goals of mesolensing observations are to discover lenses and,
for known lenses, to determine the lens mass.
We may also be able to refine measurements of the distance to and angular
speed of the lens, and also to study its multiplicity 
(Di\,~Stefano 2005a). 
Mesolensing effects are measurable in both the 
time and space domains. 
The angular resolution and photometric
sensitivity required to optimize the observing strategy will
depend on the characteristics of the background.

Time 
  variability can be most easily studied when it produces well-defined
events. I have demonstrated that the highest magnification portion
of such events is likely to be due to the lensing of a single
source and to therefore produce a light  
curve profile similar to those found in microlensing. When the source
density
is high (relative to the size of $b_i\, \theta_E$), 
the light curve shape near regions of peak
 magnification can be influenced by the presence of a small number 
of individual sources. Even then, fits to a simple
model should be possible. In the high-density regime,
 the long-time light curve
can be fit by a sequence of events, where the baseline flux may
be different at different times, because the lens is
passing across different regions of the source field.  In low-density regions,
most events will have the same characteristics as microlensing 
events and, although they can repeat, the time between events
can be tens or hundreds of years. Therefore,
even an event due to a nearby dwarf star or stellar remnant
may appear to be a ``garden variety" microlensing event.
Nevertheless, nearby lenses are more likely to be directly detectable 
in some waveband and parallax effects may also be
discernible. 

Spatial variability can provide independent checks of lensing,
and can also be used to discover lensing. Spatial variability
may be observable due to both the effects of lens motion
and to lensing induced shifts in image position as the lens
traverses the source plane. 
The sequential 
appearance and disappearance of weak-lensing perturbations
along the lens track  
is a possible signature that provides independent information about
the size of the Einstein ring.

Overall, the possible combination of 
both time and spatial effects provides more information with which
to test lensing models than is typically available for microlensing events. 
This can be so even in cases in which 
 direct imaging and spectral studies of the lens are not possible.

\subsection{Prospects}

This paper has introduced the idea of high-probability lensing, or
mesolensing. 
While many of the ideas represent natural 
extensions of 
microlensing and macrolensing studies, it is worthwhile to 
develop a framework specifically suited to high-probability lensing.
To use such lensing as an effective tool in astronomy,
computations, selection procedures, and even observing
procedures need to be designed to study the related effects.  
High-probability suggests directions for new investigations.

Computation of mesolensing effects take into account the presence of
multiple sources that may be lensed by a single mass, either simultaneously
or in sequence. The effects can be searched for in the data collected
by existing or completed monitoring programs. Modifications of the
software to include as candidates monochromatic signals, multiple-source
signals, and sequential events with possible jitter in the baseline
are needed to optimize the identification of mesolensing candidates.
Searches designed specifically to study possible spatial
effects may also be important.

Because the effects of a bright background are to require a closer
distance of approach for detectability, a subset of mesolensing
events are expected to be of short duration (a day or shorter).
In addition, fast angular velocities associated with any masses that
may be within roughly a pc of the Earth could produce even
shorter events. Observing programs that carry out occasional
high-frequency monitoring may therefore 
prove to be useful. 
Whatever the cadence of the observations, 
wide-field monitoring programs
designed to come online within the next few years
will become important pillars of lensing research.  

Many events due to high-probability lenses, especially 
those occurring against low-density backgrounds
such as the Magellanic Clouds or the Bulge, will have the same
properties as microlensing events. Others, perhaps the majority,
will be identified 
through loosening the conditions used to
select microlensing events.
In either case, to establish that the event was produced by a 
nearby lens requires additional investigation.
This includes fits to models with multiple sources,
sequential events fit to lensing models, observation
of spatial effects, and perhaps most important,   
multiwavelength  ``follow up" extending
from the infrared through the $\gamma$-ray.
Multiwavelength follow-up should certainly include correlations with
known objects and events in existing catalogs. (For example,
some NSs may have experienced episodes of $\gamma$-ray activity.)
New observations may be required to determine whether 
candidate events can be associated with nearby objects.

The most obvious application of mesolensing is to the study
of nearby stars and planets.
A much broader range of applications are, however, possible. 
For example, different 
sources,
such as nebulae and X-ray sources will be well suited for some studies.
Perhaps of most immediate interest, is the extension to lens systems that 
are intrinsically luminous. If light from a nearby star can be
blocked, using techniques similar to those planned for attempts at
imaging nearby planets, the mass of the star, and the projected separation of and
masses of any companions are potentially measurable.
 
The systematic study of lensing by nearby objects will
increase our knowledge of the mass content and characteristics
of the region of the Galaxy inhabited by the solar system.
We will learn about the masses, multiplicities, and spatial motions 
of our neighbors. The information obtained about significant numbers
of individual dark and dim nearby stellar remnants and dwarf stars 
can be woven together to 
increase our knowledge of star formation,  stellar evolution,
 and of the fundamental properties 
of each class of lens.

\bigskip

\bigskip
{\footnotesize     

%\noindent Adams, A.W., \& Bloom, J.S. 2004, astro-ph/0405266

%\noindent Adams, M.~T.~\& 
%Boroson, T.~A.\ 1979, \nat, 282, 183 

%\noindent
%Afonso, C., et al.\ 
%2003, A\&A, 404, 145 
%Bulge results

%\noindent Afonso, C., et al.\ 
%2000, \apj, 532, 340 
%binary lens SMC-1 

%\noindent
%Alcock, C., et al.\ 
%2001, \apj, 552, 259 
%binary source (LMC) + follow-up 

%\noindent
%Alcock, C., et al.\ 
%2001, \apj, 562, 337 
%Astrometry with the MACHO Data Archive. 
%I. High Proper Motion Stars toward the Galactic Bulge and Magellanic Clouds

%\noindent Alcock, C., et al.\ 
%2000, \apj, 542, 281 
%5.7 years

%\noindent
%Alcock, C., et al.\ 
%1998, \apjl, 499, L9 
%EROS and MACHO Combined Limits on Planetary-Mass Dark Matter in the Galactic Halo

%\noindent
%Alcock, C., et al.\ 
%1995, \apjl, 454, L125 
%1st detection of parallax

%\noindent Arnaboldi, M.\ 2003, Memorie 
%della Societa Astronomica Italiana Supplement, 3, 184 

%\noindent
%Arnaboldi, M. et al.\ 2004, \apjl, 614, L33  

%\noindent
%Beers, T.~C., et al.\ 
%2004, IAU Symposium, 220, 195 
%mass of MW 

%\noindent
%Belokurov, V., et 
%al.\ 2005, \mnras, 357, 17 
%The POINT-AGAPE survey - II. 
%An unrestricted search for microlensing events towards M31

\noindent
Bennett, D.~P., et al.\ 
2002, \apj, 579, 639 

\noindent
Chambers, 
K.~C., \& Pan-STARRS 2004, American Astronomical Society Meeting Abstracts, 
205,

\noindent
de Jong, J. T. A., et al. 2004, A\& A, 417, 46
%First microlensing candidates from the MEGA survey of M 31

\noindent
Di\thinspace Stefano 2005a, {\it Binary and Planet Mesolenses}, in preparation 

\noindent
Di\thinspace Stefano 2005b, {\it Mesolensing Explorations of Nearby Masses:
From Planets to Black Holes}, submitted to ApJ    

%\noindent
%Di\thinspace Stefano 2005c, {\it Lensing Tests for Supermassive 
%Black Holes}, in preparation 

\noindent
Dominik, M., \& Sahu, 
K.~C.\ 2000, \apj, 534, 213 

\noindent 
Drake, A.~J., Cook, 
K.~H., \& Keller, S.~C.\ 2004, ApJL, 607, L29 

\noindent
Dyson, F.W., Eddington, A.S., \& Davidson, C. 1920, Philos. Trans. R. Soc.\,
London, A220, 291

\noindent
Eddington, A.~S., 
Jeans, J.~H., Lodge, O.~S., Larmor, J.~S., Silberstein, L., Lindemann, 
F.~A., \& Jeffreys, H.\ 1919, MNRAS, 80, 96 

%\noindent
%Feldmeier, J.~J. et al\ 2004a, \apj, 615, 196 

%\noindent Feldmeier, J.~J.\ 2004b, 
% astro-ph/0407625 

%\noindent
%Freitag, M., Atakan 
%G{\" u}rkan, M., \& Rasio, F.~A.\ 2005, 
%astro-ph/0503130 
%Runaway collisions in young star clusters. II. Numerical results

%\noindent
%Gates, E.~I., Gyuk, G., 
%\& Turner, M.~S.\ 1995, \apjl, 449, L123 
%The Local Halo Density 

%\noindent
%Gaudi, S. \& Bloom, J. 2005.

\noindent
Gould, A., Bennett, 
D.~P., \& Alves, D.~R.\ 2004, ApJ, 614, 404 
LMC-5 

\noindent Gould, A.\ 2004, ApJ, 606, 319 
LMC-5  

\noindent Gould, A.\ 1996, \apj, 470, 201 
%theory of pixel lensing

%\noindent
%%Griest, K.\ 1991, ApJ, 366, 
412 

%\noindent
%Hogg, D.~W., Quinlan, 
%G.~D., \& Tremaine, S.\ 1991, \aj, 101, 2274 

\noindent
Holberg, J.~B., Oswalt, 
T.~D., \& Sion, E.~M.\ 2002, \apj, 571, 512 

%\noindent
%Islam, R.~R., Taylor, 
%J.~E., \& Silk, J.\ 2004, \mnras, 354, 629 
%Massive black hole remnants of the first stars - 
%III. Observational signatures from the past

\noindent
Jiang, G., et al.\ 2004, 
ApJ, 617, 1307 
%OGLE-2003-BLG-238: Microlensing Mass Estimate of an Isolated Star 

\noindent
Kochanek, C.~S.\ 2004, ApJ, 
605, 58 

\noindent
Kaspi, V.M., Roberts, M.S.E., Harding, A.K.  
2004, to appear in "Compact Stellar X-ray Sources", eds. W.H.G. Lewin and M. van der Klis,
Cambridge University Press,
astro-ph/0402136.

\noindent
Kawka, A., Vennes, S., \& 
Thorstensen, J.~R.\ 2004, \aj, 127, 1702 
Observations of White Dwarfs in the Solar Neighborhood

%\noindent
%Kerins, E., et al.\ 
%2003, \apj, 598, 993 
%Theory of Pixel Lensing toward M31. II. 
%the velocity Anisotropy and Flattening of the MACHO Distribution

\noindent
Kleinman, S.~J., et 
al.\ 2004, \apj, 607, 426 

\noindent
Liebert, J., Bergeron, 
P., \& Holberg, J.~B.\ 2005, \apjs, 156, 47 

%\noindent Luhman, K.~L., Fazio, 
%G., Megeath, T., Hartmann, L., \& Calvet, N.\ 2005, Memorie della Societa 
%Astronomica Italiana, 76, 285 

%\noindent
%Luhman, K.~L.\ 2004, \apj, 616, 
%1033 

\noindent
Luyten, W.~J.\ 1999, VizieR 
Online Data Catalog, 3070, 0 

\noindent
Macri, L. M., et al. 2001, ApJ, 549, 721

\noindent
McClintock, J.E. \& Remillard, R.A. 2003, 
to appear in "Compact Stellar X-ray Sources," eds. W.H.G. Lewin and M. van der Klis, 
Cambridge University Press, astro-ph/0306213 

\noindent
McCook, G.~P., \& Sion, 
E.~M.\ 1999, ApJS, 121, 1 

\noindent
Mao, S., et al.\ 2002, 
MNRAS, 329, 349 

\noindent
Mei, S., et al.\ 2005, 
ApJS, 156, 113 

%\noindent
%Miller, M. C., \& Colbert, E. J. M. 2004, Int. J. Mod. Phys. D, 13, 1
%Intermediate-Mass Black Holes 

%\noindent
%Mushotzky, R.\ 2004, 
%Progress of Theoretical Physics Supplement, 155, 27 
%Ultra-Luminous Sources in Nearby Galaxies 

\noindent
Nale{\. z}yty, 
M., \& Madej, J.\ 2004, A\&A, 420, 507 

\noindent
Nguyen, H.~T., 
Kallivayalil, N., Werner, M.~W., Alcock, C., Patten, B.~M., \& Stern, D.\ 
2004, ApJS, 154, 266 

%\noindent
% Okamura, S., et al.\ 
%2002, \pasj, 54, 883 

\noindent
Paczy\'nski, B.\ 1996, ARAA, 
34, 419 

\noindent
Paczy\'nski, B.\ 1986, ApJ, 
304, 1

\noindent
Paulin-Henriksson, S., et al.\ 2003, \aap, 405, 15 

%\noindent
%Portegies 
%Zwart, S.~F., Baumgardt, H., Hut, P., Makino, J., \& McMillan, S.~L.~W.\ 
%2004, \nat, 428, 724 
%Formation of massive black holes through 
%runaway collisions in dense young star clusters

%\noindent
%Reid, I.~N., Gizis, J.~E., 
%\& Hawley, S.~L.\ 2002, AJ, 124, 2721 

\noindent
Protopapas, P. et al.\, 2005, in preparation.

%\noindent
%Sakamoto, T., Chiba, 
%M., \& Beers, T.~C.\ 2003, A\&A, 397, 899 
%mass of the MW 

\noindent
Saslaw, W.~C., 
Narasimha, D., \& Chitre, S.~M.\ 1985, \apj, 292, 348 
%The gravitational lens as an astronomical diagnostic 

%\noindent
%Smith, M.~C., Mao, S., \& 
%Wo{\' z}niak, P.\ 2003, \apjl, 585, L65 

\noindent
Springel, V, White, M., \& Hernquist, L 2001, astro-ph0008144v4 

\noindent
Stubbs, C.~W., Sweeney, 
D., Tyson, J.~A., \& LSST 2004, American Astronomical Society Meeting 
Abstracts, 205,

\noindent
Tonry, J. L., \& Schneider, D. P. 1988, AJ, 96, 807

%\noindent
%Tully, R.~B.\ 1988, Journal of 
%the British Astronomical Association, 98, 316 

\noindent
 Turner, E.~L., Wardle, 
M.~J., \& Schneider, D.~P.\ 1990, \aj, 100, 146 
%A technique for detecting massive collapsed objects in the dark halo of the Galaxy

\noindent
Udalski et al, 1994, Ap J 436, 103
%OGLE 7 

\noindent
Uglesich, R.~R., 
Crotts, A.~P.~S., Baltz, E.~A., de Jong, J., Boyle, R.~P., \& Corbally, 
C.~J.\ 2004, \apj, 612, 877 
%Evidence of Halo Microlensing in M31 

\noindent
Volonteri, M., Madau, 
P., \& Haardt, F.\ 2003, \apj, 593, 661 
%The Formation of Galaxy Stellar Cores by the 
%Hierarchical Merging of Supermassive Black Holes

\noindent
Walsh, D., 
Carswell, R.~F., \& Weymann, R.~J.\ 1979, \nat, 279, 381 

}
\begin{table*}
\centering{
\caption{Probability of Multiple-Source Events}
\begin{tabular}{|c||c|c|ccccc|}
\hline
$\theta_{b}/\theta_1$ & duty cycle & $N_{tot}$ & $P_1$ & $P_2$ & $P_3$ & $P_
4$ & $P_5$\cr
\hline
    0.05 &     0.002 &     313 &     0.99 &     0.00 &     0.00 &     0.00 &
     0.00  \cr
    0.10 &     0.011 &     668 &     0.93 &     0.06 &     0.01 &     0.00 &
     0.00  \cr
    0.15 &     0.026 &    1023 &     0.82 &     0.12 &     0.03 &     0.01 &
     0.01  \cr
    0.20 &     0.054 &    1465 &     0.66 &     0.15 &     0.06 &     0.04 &
     0.02  \cr
    0.25 &     0.108 &    2110 &     0.45 &     0.16 &     0.07 &     0.07 &
     0.07  \cr
    0.30 &     0.173 &    2621 &     0.34 &     0.14 &     0.10 &     0.10 &
     0.14  \cr
\hline
\end{tabular}
\par
\medskip
%\footnotesize
Notes:
An ``event"
starts when the lens comes within $\theta_b$ of
any source and continues as long as the lens is within $\theta_b$ of
any source; thus events include perturbations due to the simultaneous
lensing of multiple stars.
The duty cycle is the
total time duration of all events occurring during the simulation,
divided by the total time of the simulation.
$N_{tot}$ is the total number of events.
$P_i$ is the fraction of all
events in which $i$ source stars were simultaneously
 magnified during the event (the projected lens
position is within $\theta_b$ of $i$ stars at some point during the event).
}
\par
\end{table*}
 
\vspace{+0.5cm}
\noindent{Acknowledgements: This work has been in development
for a long time and there are many people to thank. First, Tsafrir
Kolatt worked with me on a related question almost a decade ago,
considering spatial and time variations due to massive BHs in the 
Halo. Thanks are due to P.L. Schechter  and E.L. Turner for
discussions of that idea and to the participants of a 1997 Aspen
workshop; especially valuable was feedback from members of the AGAPE team,
who tested feasibility. Calculations that led to the present
conceptualization were begun at IUCAA in 1998. Jason Li,
working under the aegis of {\it RSI} program run by the 
{\it Center for Excellence in Education}, 
 participated
in checks of this work in 2003 and developed the idea of creating
a catalog of high-probability lenses. I especially want to thank
Eric Pfahl for his encouragement to pursue this line of investigation,
and for technical discussions. Discussions with many others
during the past year have also been helpful, especially 
Charles Alcock, Lorenzo Faccioli, Arti Garg, 
B. Scott Gaudi,  Matthew J. Holman,
Nitya Kallivayalil, 
Kevin L. Luhman, Christopher Night, Brian M. Patten, 
and Christopher Stubbs.
This work was supported in part by NAG5-10705.        
}

%\vspace{-1 true in}
\begin{figure*}
%\vspace{-1 true in}
\psfig{file=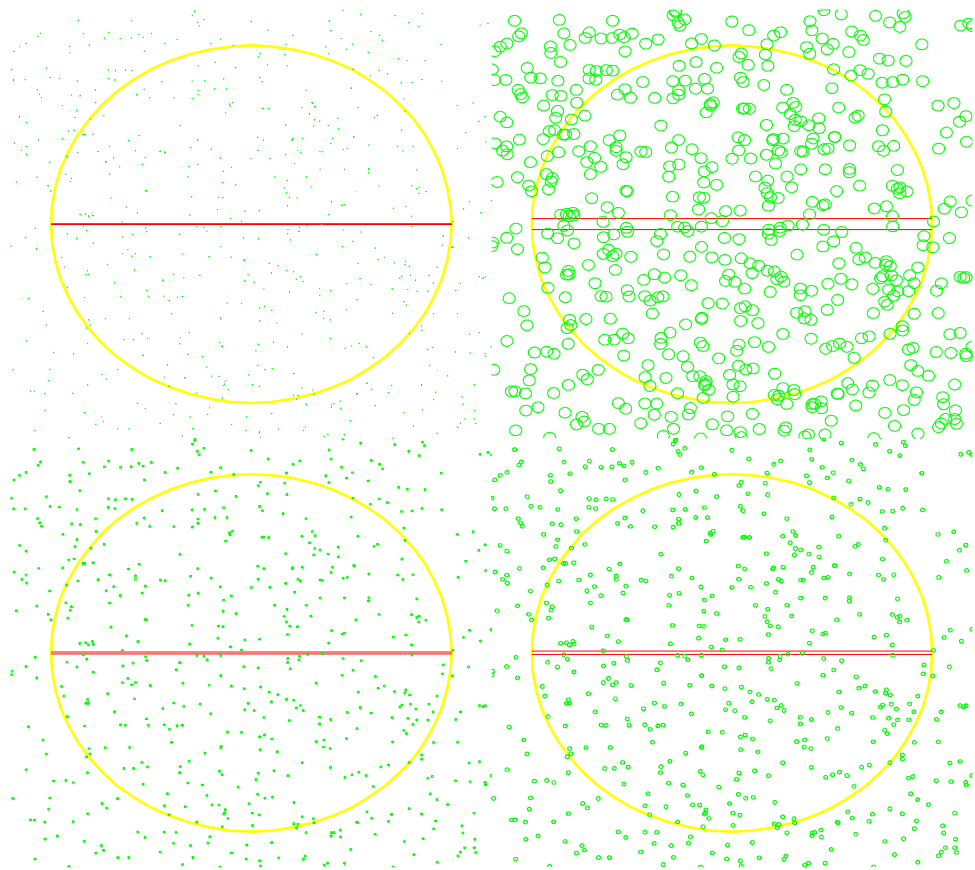,height=6in,width=6in,angle=-0.0}
\vspace{1 true in}
\caption{
The position of each circle represents the randomly-generated
position of a source star.
The radius of the circle is equal to $\theta_b,$ and is
the same for all source stars.  
In these $4$ panels, the physical density of source stars has just
one value:  $1$ star$/\theta_E^2,$ corresponding to $\theta_1=0.56\, \theta_E.$
Yet the panels span the range from low density (upper left), counter clockwise to
high density (upper right), because the value of $\theta_b$ changes from
panel to panel. Starting at the upper left  panel and
moving counter clockwise, $\theta_b = 0.01, 0.05, 0.1, 0.3.$ 
The large yellow circle has a radius of $10\, \theta_E.$
the red slice through the center of each panel represents 
is $2\, \theta_b$ wide and represents the possible track of a lens
and the region around it in which source stars can be detectable lensed. 
If $\omega \sim 20\, \theta_E$ per year, then each red track
shows the distance traversed by the lens in a time of approximately
one year.  
For a more realistic luminosity function, there would be a large
range of radii, $\theta_b.$
Note that in the low-density regime, each event is caused by the lensing
of a single source, and that events are well separated; repeats are
rare. In the high-density regime, multiple-source events are 
common and can be of long duration; ``repeats" are common and the duty cycle
is high.
}
\end{figure*}

\vspace{.3 true in}
\begin{figure*}
\psfig{file=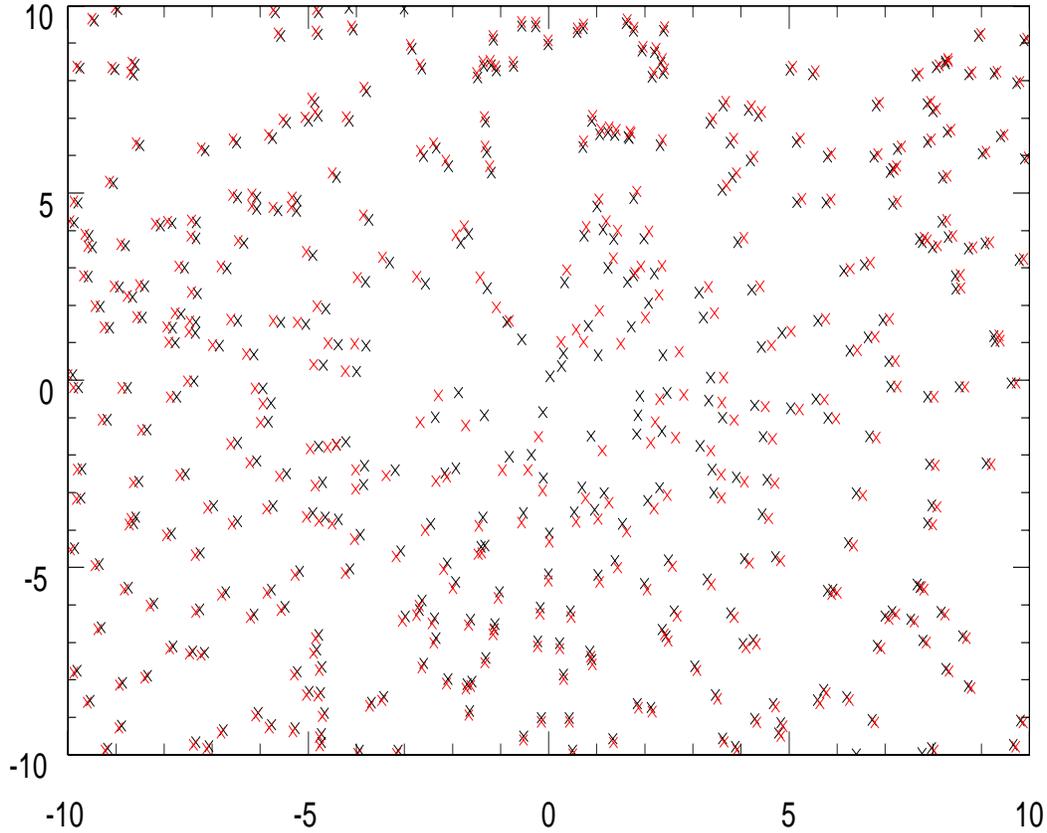,height=8in,width=6in,angle=-0.0}
\vspace{-3 true in}
\caption{
Shifts in image position due to a lens located at (0,0);
$\theta_E$ is the unit of distance along both the vertical
and horizontal axes.
Black crosses: randomly generated positions of source stars;
the source density is 1 per $\theta_E^2$.
Red crosses: positions of positive parity images.
}
\end{figure*}

\end{document}